\documentclass[useAMS,usenatbib]{raa}
\usepackage{graphicx}

\begin{document}

\title{Orbital X-ray modulation study of three Supergiant HMXBs}

\volnopage{ {\bf 2009} Vol.\ {\bf 9} No. {\bf XX}, 000--000}
\setcounter{page}{1}

\author{Chetana Jain
      \inst{1,2}
   \and Biswajit Paul
      \inst{2}
   \and Anjan Dutta
      \inst{1}
   }

\institute{Department of Physics and Astrophysics, University of Delhi,  Delhi 110007, India; {\it chetanajain11@gmail.com}\\
	        \and
             Raman Research Institute, Sadashivnagar, C. V. Raman Avenue, Bangalore 560080, India}

\abstract{We present the orbital X-ray modulation study of three high mass X-ray binary systems, 
IGR J18027$-$2016, IGR J18483$-$0311 and IGR J16318$-$4848 using data obtained with $RXTE$-ASM, 
$Swift$-BAT and $INTEGRAL$-ISGRI. Using the long term light curves of the eclipsing HMXB IGR J18027$-$2016, 
obtained with $Swift$-BAT in the energy range 15$-$50 keV and $INTEGRAL$-ISGRI in the energy range 22$-$40 
keV, we have determined three new mid eclipse times. The newly determined mid eclipse times together with 
the known values were used to derive an accurate value of the orbital period of 4.5693(4) d at MJD 52168 
and an upper limit of 3.9(1.2)$\times$10$^{-7}$ d d$^{-1}$ on the period derivative. We have also 
accurately determined an orbital period of 18.5482(88) d for the intermediate system IGR J18483$-$0311, 
which displays an unusual behaviour and shares many properties with the known SFXTs and persistent 
supergiant systems. This is a transient source and the outbursts occur intermittently at intervals 
of 18.55 d. Similarly, in the third supergiant system, IGR J16318$-$4848, we have found that the outbursts 
are separated by intervals of 80 d or its multiples, suggesting a possible orbital period.
	 }

\authorrunning{C. Jain, B. Paul \& A. Dutta}
\titlerunning{Orbital X-ray modulation study of three Supergiant HMXBs}
\maketitle
	 
\section{Introduction}

The INTEernational Gamma-Ray Astrophysics Laboratory, $INTEGRAL$ was launched in 2002 
October (Winkler et al. 2003) and has discovered many new hard X-ray sources during the regular 
survey of the Galactic center (Revnivtsev et al. 2004, Bird et al. 2007, Kuulkers et al. 2007). 
In the pre-$INTEGRAL$ era, most of the known HMXBs were Be-X-ray binary systems, but the $INTEGRAL$ 
observations have significantly changed the statistics concerning the nature of the companion star 
of HMXBs. For instance, Liu et al. (2000) had mentioned 54 Be X-ray systems and 7 supergiant X-ray 
binary systems in their catalog of HMXBs. But, due to a large field of view of the instruments 
on board $INTEGRAL$ (Lebrun et al. 2003, Ubertini et al. 2003), and a high sensitivity at hard X-rays, 
several new HMXBs have been discovered and the proportion of supergiant systems has increased 
drastically. It has particularly revealed many new HMXBs which are obscured by the dense and highly 
absorbing circumstellar wind of the companion, because of which these X-ray sources are not observable 
at low energies. Bird et al. (2007) identified 68 HMXBs in their third IBIS/ISGRI soft $\gamma$-ray 
survey catalog. Out of these, 24 systems were identified as Be X-ray systems and 19 as supergiants. 
In about 5 years since its launch, $INTEGRAL$ has revealed two distinct classes of supergiant X-ray 
binary systems. The first class includes obscured persistent sources (Kuulkers 2005) and the second 
class includes sources displaying a short transitory nature (Supergiant Fast X-ray Transients, SFXTs) 
with outbursts lasting for a few hours (Negueruela et al. 2006, Sguera et al. 2005, 2006). \textbf{Further, 
several persistent low luminosity, slow X-ray pulsators have also been identified, some of 
which belong to HMXB systems (Kaur et al. 2009).} 

\begin{figure}
\centering
\includegraphics[height=3.5in, width=3in, angle=-90]{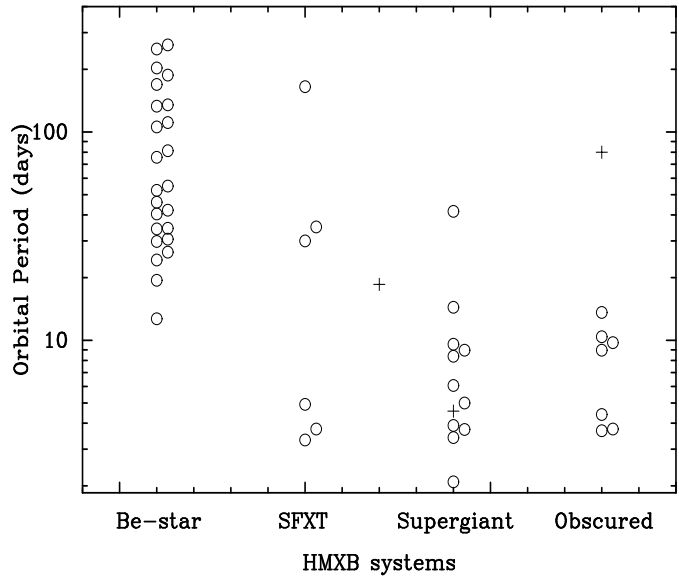}
\caption{Orbital period distribution in high mass X-ray binaries categorized into Be-star system, 
the SFXTs, the persistent supergiants and the obscurred sources. The $+$ indicate the orbital period 
of the three systems studied in the present work. \label{fig:fig1}}
\end{figure}

\textbf{Figure~\ref{fig:fig1} shows the orbital period distribution of the different sub-classes of high 
mass X-ray binaries. We have categorized the distribution into Be-star systems, the SFXTs, the persistent 
supergiant systems and the obscurred systems. Orbital period of only those HMXBs are shown which are 
mentioned in the HMXB catalogue by Liu et al. (2007). It is clear from the figure that the orbital period 
in Be-star systems range from 12$-$262 d, whereas the orbital period of supergiant systems are relatively 
shorter. The orbital period in these systems are mostly less than 15 d, except for one system having an 
orbital period of 42 d. Amongst the SFXTs listed in the catalogue, the orbital period is known in six 
systems and it varies over a wide range of 3.3$-$165 d. The obscurred sources tend to have small orbital 
periods with the widest known orbital period of about 13 d. The orbital periods of the three sources 
studied in this work are marked with a $``$plus'' sign. The first source is an intermediate SFXT having 
an orbital period of about 18 d. The second source is a supergiant HMXB, whereas the third one is a 
highly obscurred system.}

We have carried out orbital modulation studies of bright $INTEGRAL$ sources and have discovered a very 
short orbital period in one source, SFXT IGR J16479$-$4514 (Jain et al. 2009). Here, we present results 
from three of the brightest $INTEGRAL$ sources, IGR J18027$-$2016, IGR J18483$-$0311 and IGR J16318$-$4848. 
These systems have a late O/early B type supergiant companion and are highly absorbed sources. While 
IGR J18027$-$2016 and IGR J18483$-$0311 are pulsars, nature of the compact object in IGR J16318$-$4848 is 
not yet known in spite of extensive observations with different observatories. IGR J18027$-$2016 is an 
eclipsing high mass X-ray binary consisting of a neutron star spinning with a period of 139 s. The eclipses 
provide a good fiducial timing marker for precise determination of the orbital evolution. IGR J18483$-$0311 
is an intermediate system whose position on the Corbet diagram (Corbet 1986) indicates that it is likely a 
Be system, but a periodic fast X-ray transient activity observed in this system is typical of an SFXT 
system. It is therefore important to determine its orbital parameters. IGR J16318$-$4848 is one of the 
most absorbed Galactic sources known with an enormously high column density. A study of orbital 
modulation in all these systems is important to understand the mechanism for the short and long duration 
outbursts and is also useful to plan future orbital phase dependent observations. 

IGR J18027$-$2016 is spatially associated with the X-ray pulsar SAX J1802.7$-$2017 (Augello et al. 2003) 
which was serendipitously discovered during a $Beppo$-SAX observation of the LMXB GX 9+1 in 2001 September. 
It is an eclipsing HMXB system and harbors an X-ray pulsar accreting matter from the stellar wind of the 
companion star, which is a late O/early B-type supergiant with a mass 
of 18.8$-$29.3 M$_{\odot}$ (Hill et al. 2005). From the $Beppo$-SAX observations, Augello et al. (2003) 
determined a pulse period of 139.612 s and from the pulse arrival time analysis, they determined an orbital 
period of $\sim$ 4.6 d. It was later confirmed by Hill et al. (2005), who determined an orbital period 
of 4.5696(9) d from the eclipse timing measurement of the ISGRI data. They also determined a projected 
semimajor axis (a$_{x}$ $\sin$ (i)) of 68 lt-s and a mass function of 16 M$_{\odot}$, from which they 
concluded the mass of the donor to be 18.8$-$29.3 M$_{\odot}$ and radius of 15.0$-$23.4 R$_{\odot}$. Spectral 
analysis with $XMM-Newton$ and $INTEGRAL$-ISGRI indicate a strong intrinsic absorption with a hydrogen column 
density N$_{H}$ of 6.8 $\times$ 10$^{22}$ cm$^{-2}$ (Hill et al. 2005). Lutovinov et al (2005) fitted 
the 18$-$60 keV spectrum with a powerlaw alongwith an exponential cutoff at high 
energies (E$_{c} \sim$ 18 keV). 
 
IGR J18483$-$0311 was discovered with $INTEGRAL$ during a survey of the Galactic plane in 2003 
April (Chernyakova et al. 2003). An average flux of $\sim$ 10 mCrab in the 15$-$40 keV was observed, 
which decreased to 5 mCrab in the 40$-$100 keV energy range. IGR J18483$-$0311 is a high mass X-ray 
binary with an early B-type supergiant companion star (Rahoui et al. 2008). From the timing analysis 
of the $RXTE$-ASM light curve, Levine et al. (2006) reported a 18.55(5) d orbital period and 21 s 
X-ray pulsations were reported by Sguera et al. (2007). The source displays an unusual behaviour and 
shares many properties with the known SFXTs and persistent supergiant systems. Association with a B0.5Ia 
supergiant companion star (Rahoui et al. 2008) and a fast X-ray transient activity (Sguera et al. 2007), 
indicate that the system could be an SFXT. But, the outbursts last for a few days, in contrast to a few 
hours long outbursts seen in other well known SFXTs (Sguera et al. 2007). The quiescent emission level 
is also higher in IGR J18483$-$0311, yielding an L$_{max}$/L$_{min}$ ratio of $\sim$ 10$^{3}$, whereas, 
in SFXTs, the ratio is 10$^{4}$ - 10$^{5}$. The system is therefore considered to be an $``$intermediate" 
SFXT. The 3$-$50 keV spectra is well fitted by an absorbed powerlaw with a photon index of 1.4 and a 
cutoff at 22 keV. A high intrinsic absorption is also seen with a column 
density, N$_{H}$ of 9 $\times$ 10$^{22}$ cm$^{-2}$. Spectra during the outbursts is well fit by an 
absorbed bremsstrahlung with N$_{H}$ of 7.5 $\times$ 10$^{22}$ cm$^{-2}$ and 
kT $\sim$ 21.5 keV (Sguera et al. 2007).   

IGR J16318$-$4848 is one of the highly obscurred Galactic X-ray sources discovered 
by $INTEGRAL$ (Courvoisier et al. 2003) and follow-up by the $XMM-Newton$ observatory accurately localized 
its position (de Plaa et al. 2003, Schartel et al. 2003). A flux of 50$-$100 mCrab was observed in 
the 15$-$40 keV energy band with a significant variability on timescales of more 
than 1000 s (Walter et al. 2003). Observations made with the $XMM-Newton$ revealed the presence of 
strong Fe-K$_{\alpha}$, Fe-K$_{\beta}$ and Ni-K$_{\alpha}$ emission 
lines (Schartel et al. 2003, de Plaa et al. 2003), alongwith a highly absorbed 
powerlaw ($\Gamma \sim$ 1.7-2.1) continuum (Matt $\&$ Guainazzi 2003). The IR spectrum is also rich in 
emission lines, various orders of H, He I and He II (Kaplan et al. 2006). IGR J16318$-$4848 is surrounded 
by dense circumstellar material and powered by accretion from a stellar 
wind (Revnivtsev et al. 2003, Filliatre $\&$ Chaty 2004). From the 
archived $ASCA$ observations, Revnivtsev et al. (2003) determined an enormously high column 
density, N$_{H}$ $\simeq$ 10$^{24}$ cm$^{-2}$, due to which, the source is not observable at 
energies below 4 keV. 

We report the timing analysis of these three bright supergiant systems, IGR J18027$-$2016, 
IGR J18483$-$0311 and IGR J16318$-$4848. Using the data obtained from $Swift$-BAT and $INTEGRAL$-ISGRI, 
we have determined the orbital periods of IGR J18027$-$2016 and  IGR J18483$-$0311. We have also 
discovered an 80 d periodicity in the occurence of outbursts in IGR J16318$-$4848, which is possibly 
indicative of a binary orbital period.

\section{Observations and analysis}
We have used data obtained with instruments on board Rossi X-ray Timing Explorer ($RXTE$), $Swift$ Gamma 
Ray Burst Explorer and INTErnational Gamma-Ray Astrophysics Laboratory ($INTEGRAL$). The three sources 
were regularly monitored by the All Sky Monitor (ASM) on board the $RXTE$. The ASM data used for the 
present work covered the time span from MJD 50088 to MJD 54860. The 15$-$50 keV light curves 
of IGR J18027$-$2016, IGR J18483$-$0311 and IGR J16318$-$4848 were obtained from the Burst Alert 
Telescope (BAT; Barthelmy et al. 2005) on board the $Swift$ observatory. The observations covered 
the time range from MJD 53413 to MJD 54867. For all the three sources, the 22$-$40 keV long 
term $INTEGRAL$-ISGRI light curve spanned $\sim$ 1350 days.  

\begin{figure}
\centering
\includegraphics[height=3.2in, width=2in, angle=-90]{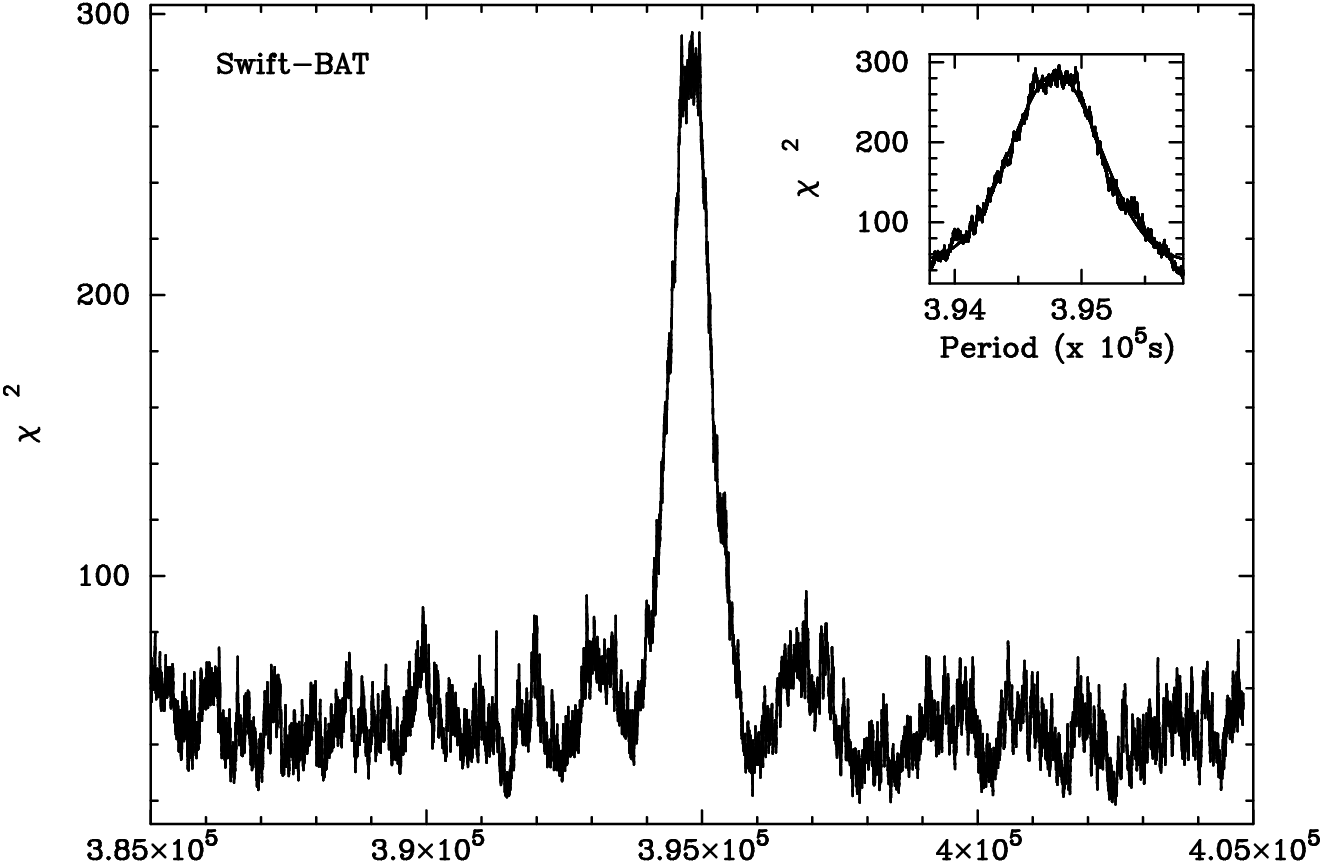}
\includegraphics[height=3.2in, width=2in, angle=-90]{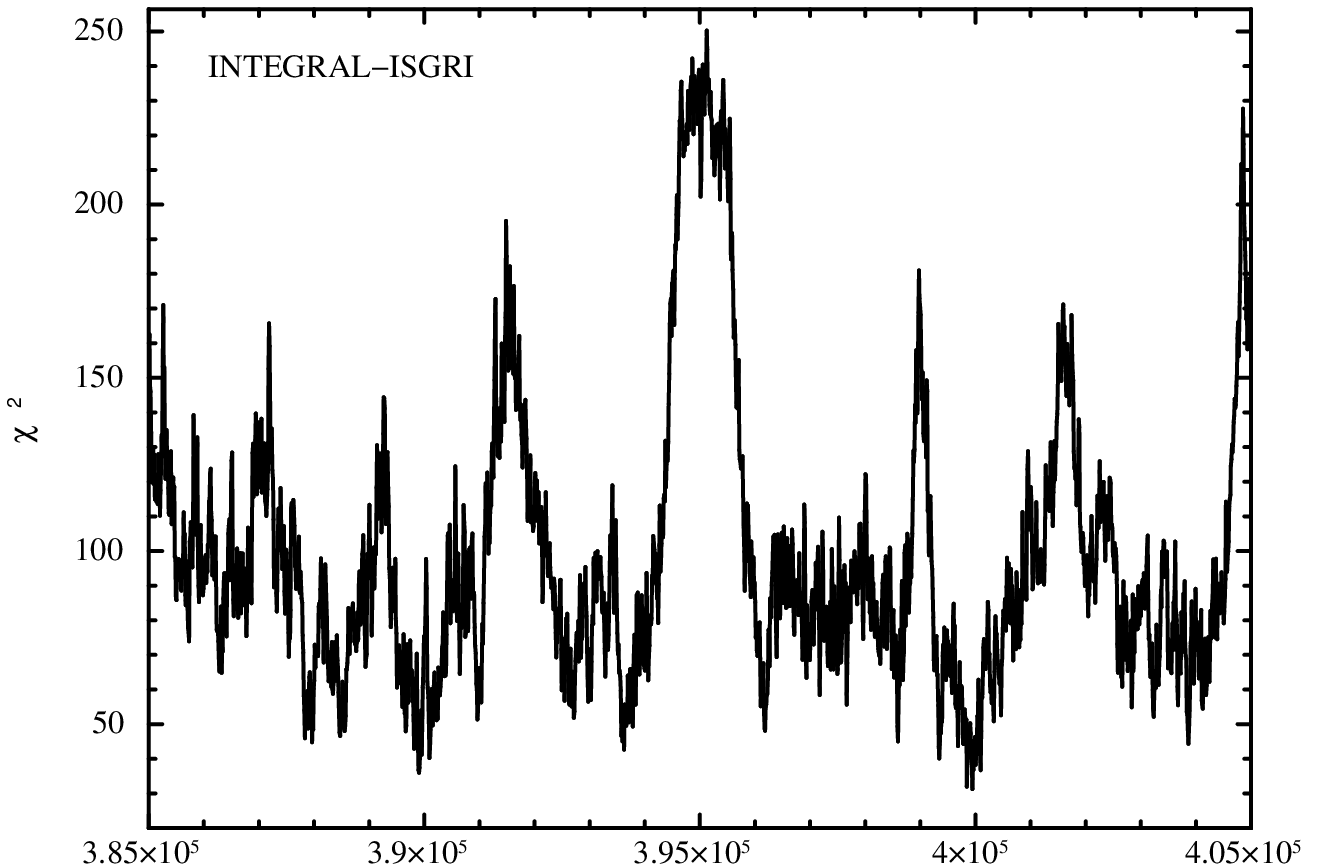}
\includegraphics[height=3.0in, width=2in, angle=-90]{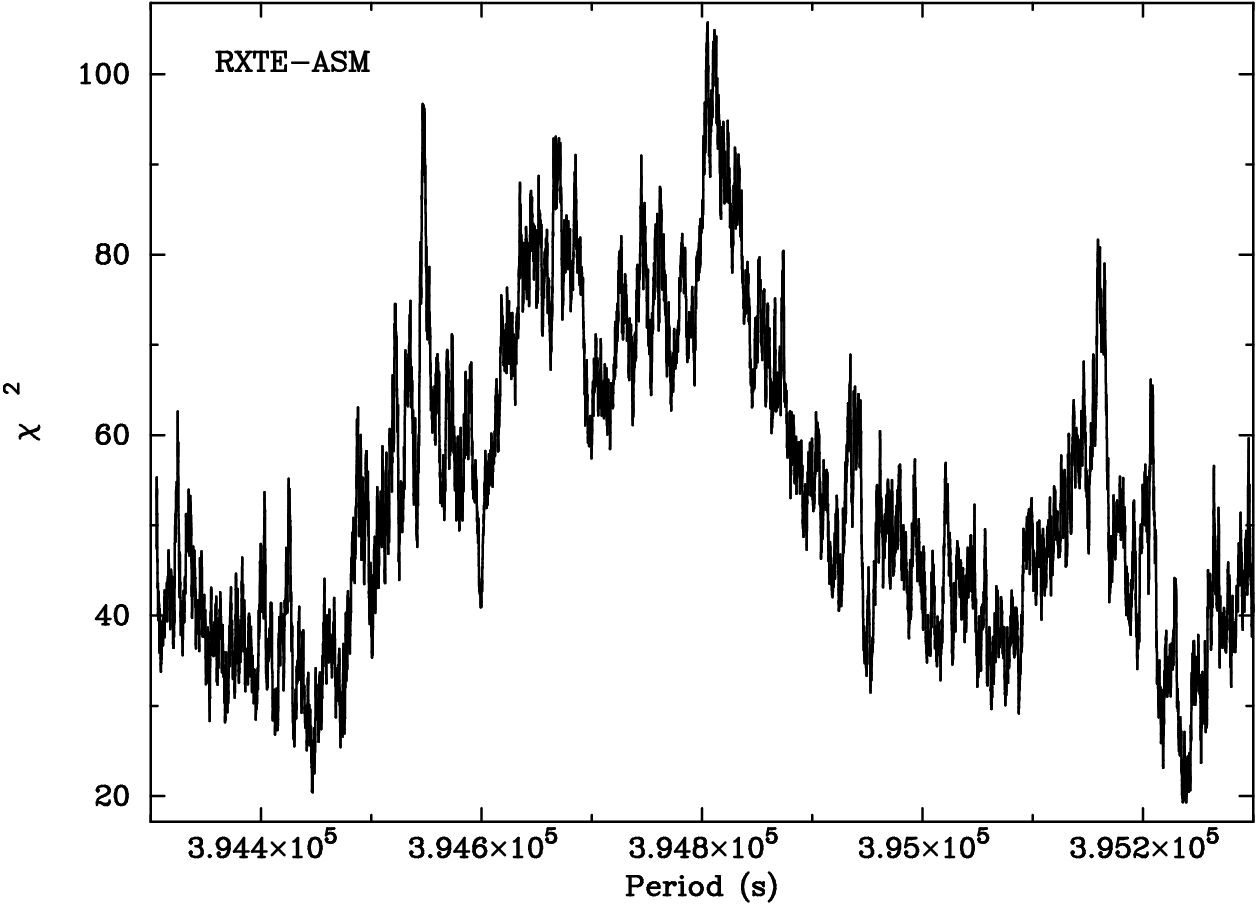}
\caption{Result from $efsearch$ on the light curve of IGR J18027$-$2016. The top panel shows the result 
from the $Swift$-BAT observations and the inset figure shows the expanded view around the peak. The 
solid line represents the best fit gaussian curve with the centre at 394793(103) s. The second and third 
panels show the $efsearch$ results from $INTEGRAL$-ISGRI and $RXTE$-ASM light curves. The peak in 
the $INTEGRAL$-ISGRI result corresponds to a period of 395056(210) s, while from the $RXTE$-ASM data, 
we determined a period of 394805(185) s. \label{fig:f1}}
\end{figure}

\textbf{IGR J18027$-$2016:} The long term $Swift$-BAT, $INTEGRAL$-ISGRI and $RXTE$-ASM light curves 
of IGR J18027$-$2016 were corrected for the earth motion using the $earth2sun$ tool of the HEASARC 
software package $``$Ftools" ver6.5.1. We searched for the orbital period using the ftool - $efsearch$, 
which folds the light curve with a large number of trial periods around an approximate period. 
Figure~\ref{fig:f1} shows the $efsearch$ result on the light curve of IGR J18027$-$2016.
The top panel shows the result of period search on the long term $Swift$-BAT light curve. The peak here 
corresponds to the periodicity in the light curve. The inset figure is the expanded view around the peak. 
A gaussian fit around the peak gave the gaussian center as 394793(103) s (4.5693(11) d). The $efsearch$ 
result of the $INTEGRAL$ data, over the same range as in $Swift$-BAT, is shown in the second panel. The 
main peak corresponds to a period of 395056(210) s (4.5723(24) d). It should be noted that the 
present $INTEGRAL$ dataset is longer than that analyzed by Hill et al. 2005, who determined a period 
of 4.570 (3) d using the ISGRI data spanning $\sim$417 days. The $INTEGRAL$ data used in the present 
work covered the time range from MJD 52698 to MJD 54041. The third panel shows the $efsearch$ result 
on the 5$-$12 keV $RXTE$-ASM light curve. A peak is present but with a poor significance. The peak 
corresponds to a period of 394805(185) s (4.5695(21) d). 

\begin{figure}
\centering
\includegraphics[height=3in, width=5in, angle=-90]{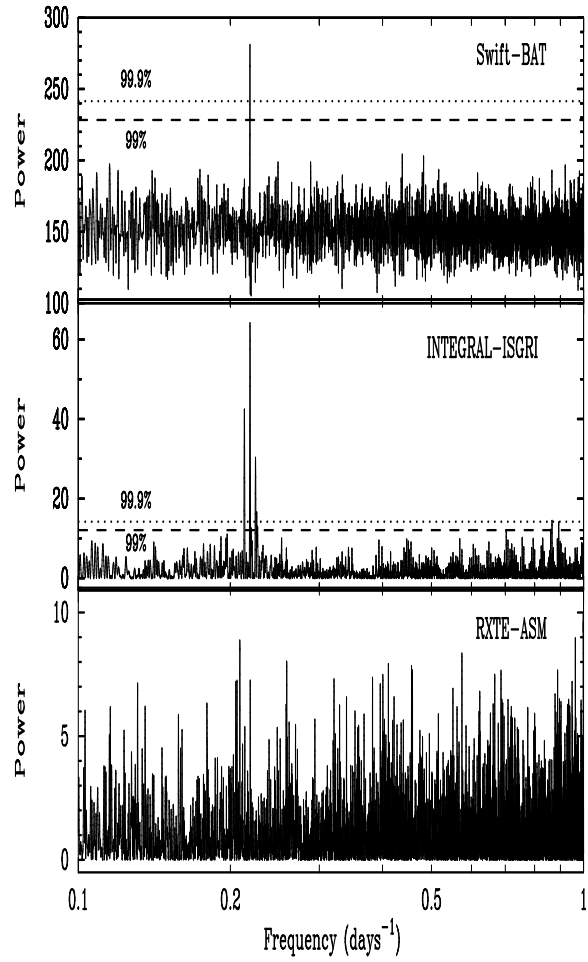}
\caption{The Lomb-Scargle periodogram generated from the $Swift$-BAT, $INTEGRAL$-ISGRI and $RXTE$-ASM 
light curves of IGR J18027$-$2016. The highest peak in the top and the middle panels correspond to a 
frequency of 0.218 d$^{-1}$, .i.e. a period of 4.5871 d. \label{fig:f2}}
\end{figure}

We have also confirmed the periodicity in the light curves of IGR J18027$-$2016, using the Lomb-Scargle 
periodogram method by means of the fast implementation of Press $\&$ Rybicki (1989) and Scargle (1982) 
technique. Figure~\ref{fig:f2} shows the periodogram generated using the $Swift$-BAT, $INTEGRAL$-ISGRI 
and $RXTE$-ASM light curves. 
As seen in Figure~\ref{fig:f2}, a clear peak is present in the periodogram generated from 
the $Swift$ and $INTEGRAL$ light curves. But, periodicity could not be confirmed from the $RXTE$-ASM 
observations. The power spectrum in the case of $Swift$-BAT and $INTEGRAL$-ISGRI data, peaks 
at 0.218 d$^{-1}$, which corresponds to a periodicity of 4.5871 d. This result is in sync with the 
values determined by the $efsearch$ analysis and those reported by Hill et al. (2005). The significance 
of these peaks was confirmed by a randomization test. For both, $Swift$-BAT and $INTEGRAL$-ISGRI light 
curves, the time stamps of the observed count rates were randomly shuffled and a periodogram was generated 
from the resulting time series. We simulated 10,000 light curves and determined the maximum power for 
both the cases. As shown in Figure~\ref{fig:f2}, the horizontal lines in the top two panels show the 
significance level. The dotted and dashed lines respectively show the 99.9$\%$ and 99$\%$ significance 
power among the randomized light curves. This imply that a peak power of 281 and 61 in the 
original $Swift$-BAT and $INTEGRAL$-ISGRI periodograms is unlikely to occur by chance and therefore 
the period detection is significant. 

\begin{figure}
\centering
\includegraphics[height=3.0in, angle=-90]{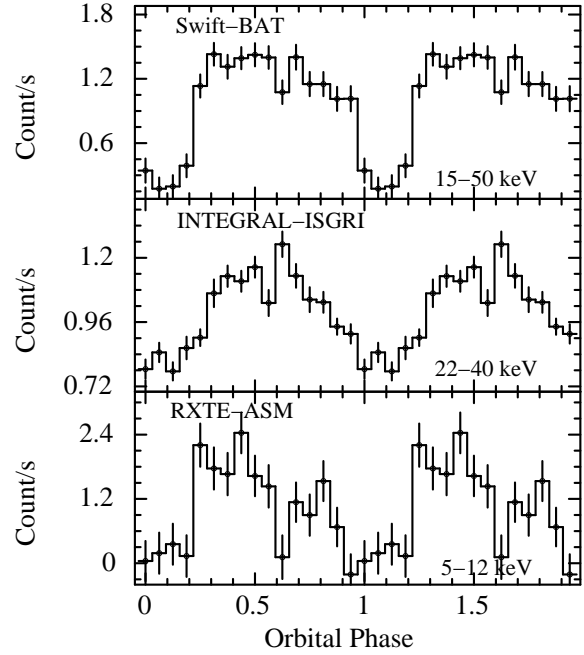}
\caption{The light curve of IGR J18027$-$2016 folded in 16 bins with a period of 394805 s. The folded 
light curves of $Swift$-BAT, $INTEGRAL$-ISGRI and $RXTE$-ASM observations are shown in the top, middle 
and bottom panels respectively. \label{fig:f3}}
\end{figure}

To determine the long term orbital solution, we folded the $Swift$-BAT, $INTEGRAL$-ISGRI and $RXTE$-ASM 
light curves in 16 phasebins with a period of 394805 s. The folded light curves are shown in Figure~\ref{fig:f3}. 
A sharp eclipse is clearly seen in the folded $Swift$-BAT light curve. The eclipse lasts for $\sim$ 0.2 
orbital phase. A clear eclipse in also seen in the folded $INTEGRAL$-ISGRI light curve but it is not sharp 
as compared to the eclipse seen in the folded $Swift$-BAT light curve. The eclipse detection in the 
folded $RXTE$-ASM light curves is not significant, but we emphasize that it occurs at the same phase as 
seen in the other two observations. We fitted a gaussian to the eclipse phase and the center of the best 
fit gaussian gives the mid eclipse time for that observation. \textbf{From the folded $Swift$-BAT light 
curve, we determined an eclipse half width of 0.1923 orbital phase. This implies an eclipse half angle 
of 0.604 radians. The mass of the companion star is known to lie within a range 
of 19$-$29 M$_{\odot}$. Therefore, assuming a canonical mass of 1.4 M$_{\odot}$ for the neutron 
star, the lower limit on the companion star radius will lie in the range 16.4$-$24.7 R$_{\odot}$.}

\begin{figure}
\centering
\includegraphics[height=3.0in, angle=-90]{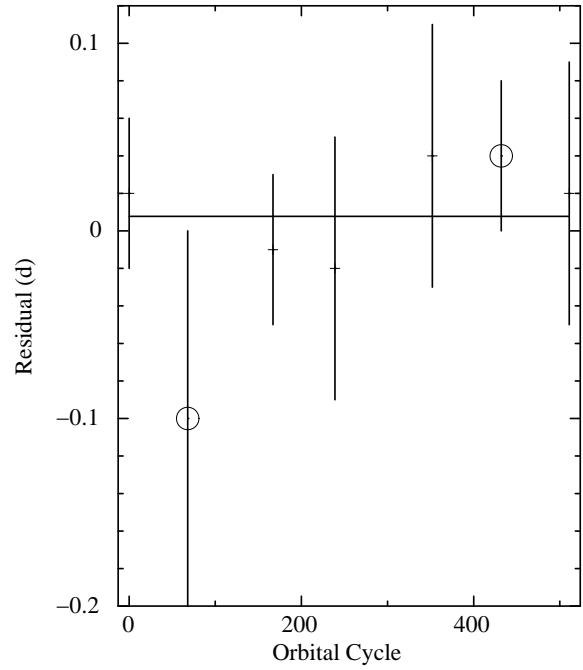}
\caption{The mid eclipse time residuals of IGR J18027$-$2016 are plotted as a function of the orbital 
cycle, relative to the best fit linear ephemeris (P$_{orb}$ = 4.5693 d at MJD 52168). The mid eclipse 
times are tabulated in Table 1. The $``\circ$" represents the $RXTE$-ASM and average $Swift$-BAT, 
respectively. \label{fig:f4}}
\end{figure}

Table 1 gives a log of the mid eclipse times determined from each observation. 
\begin{table}
\caption{X-ray mid-eclipse times of IGR J18027$-$2016.}
\begin{center}
\begin{tabular}{c c c l} 
\hline\hline
Cycle & Mid eclipse time & Uncertainty & Satellite\\
      & (MJD)		 & (d)	       &	\\
\hline
 0$\dag$    & 52168.26  & 0.04& $Beppo$-SAX\\
 68   	    & 52478.78  & 0.12& $RXTE$-ASM\\
 167$\dag$  & 52931.37  & 0.04& $INTEGRAL$-ISGRI\\
 239  	    & 53260.37  & 0.07& $INTEGRAL$-ISGRI\\
 352  	    & 53776.82  & 0.07& $Swift$-BAT\\
 511  	    & 54503.38  & 0.07& $Swift$-BAT\\
\hline
\end{tabular}
\end{center}
$\dag$Reported by Hill et al. 2005
\label{Table 1}  
\end{table}
We determined two mid eclipse times from the $Swift$-BAT data and one from $INTEGRAL$-ISGRI data. The 
eclipse seen in the $RXTE$-ASM light curve is not sharp and hence the determination of mid eclipse time 
involves a large error. We then combined these newly determined mid eclipse times with the known values 
and fitted a quadratic model to the ephemeris history. We determined an orbital period (P$_{orb}$) 
of 394787(34) s (4.5693(4) d) and a period derivative of 3.9(1.2)$\times$10$^{-7}$ d d$^{-1}$ at MJD 52168. 
We then subtracted the best fit linear component from the ephemeris history and the residual is plotted 
in Figure~\ref{fig:f4}. There are only few mid eclipse times reported for this source, therefore, it is 
not possible to accurately determine the orbital evolution of this binary system. But it should be noted 
that the period is indeed evolving and probably, future observations of the source can lead to 
determination of the orbital evolution in this system. \textbf{In particular, since this is an eclipsing 
system, the optical measurements of the companion star can be useful to place a constraint on the rate of 
mass loss from the donor star.}

\begin{figure}
\centering
\includegraphics[height=3.2in,  width=6.0in, angle=-90]{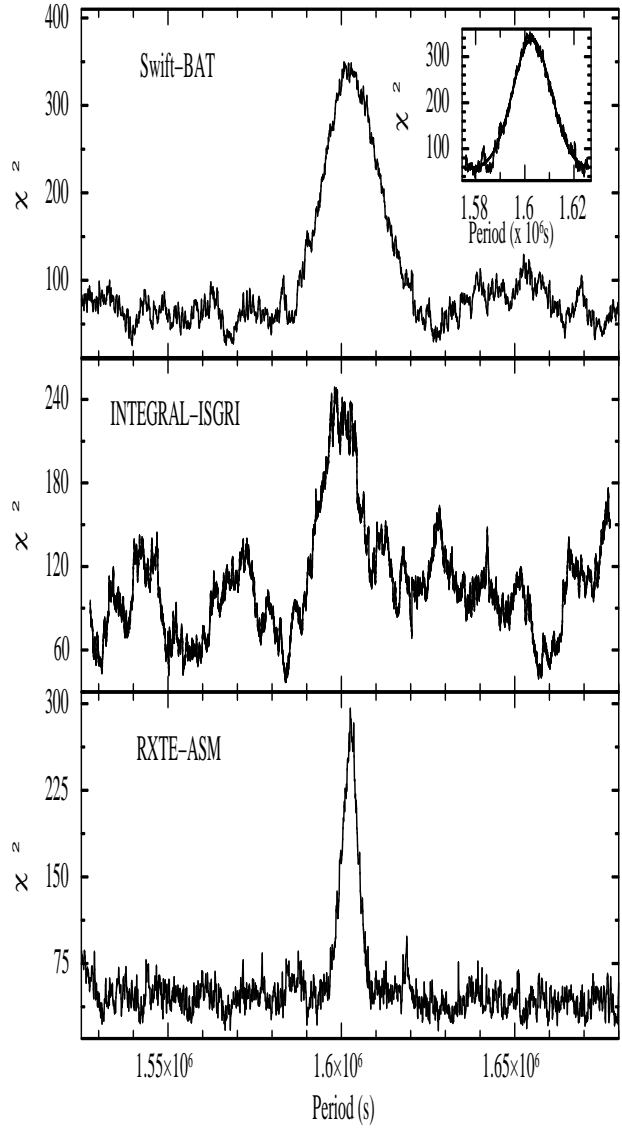}
\caption{Results from $efsearch$ on the light curve of IGR J18483$-$0311. The top, middle and bottom 
panels show the result from the $Swift$-BAT, $INTEGRAL$-ISGRI and $RXTE$-ASM observations, respectively. 
The inset figure in the top panel shows the expanded view around the peak determined from the $Swift$-BAT 
observations. The solid line represents the best fit gaussian curve with the centre at 18.550(26) d. 
The peaks in the $INTEGRAL$-ISGRI and $RXTE$-ASM results respectively correspond to a period 
of 18.521(34) d and 18.5482(88) d. \label{fig:f5}}
\end{figure}

\textbf{IGR J18483$-$0311:} The $efsearch$ period search result on the long term $Swift$-BAT, 
$INTEGRAL$-ISGRI  and $RXTE$-ASM light curves of IGR J18483$-$0311 are shown in Figure~\ref{fig:f5}. 
Clear peaks are seen in all the three results. A gaussian was fit around the peak in the $Swift$-BAT 
period search results (inset figure in the top panel) and the peak center determined. An orbital 
period of 1602796(2268) s (18.550(26) d) was determined. The peak in the $efsearch$ result of 
the 22-40 keV $INTEGRAL$-ISGRI and 5$-$12 keV $RXTE$-ASM light curves corresponds 
to 1600227(2989) s (18.521(34) d) and 1602571(767) s (18.5482(88) d), respectively. These results are 
an improvement over the results obtained by Sguera et al. (2007), who analyzed $\sim$ 1142 days of data 
from the $INTEGRAL$ observations and determined an orbital period of 18.52 d. Whereas, the present result 
is more complete with the $INTEGRAL$ data  covering the time range from MJD 52704 to MJD 54053. 

IGR J18483-0311 is a bursting source and therefore, initially we could not determine an accurate period 
from the $efsearch$ analysis of the entire $INTEGRAL$ light curve. But after the removal of 5$\sigma$ 
bursts (as explained later), we determined the orbital period accurately. We have confirmed the 
periodicity in the long term $Swift$-BAT and $RXTE$-ASM light curves by using the Lomb-Scargle 
periodogram technique as mentioned in the case of IGR J18027$-$2016.
\begin{figure}
\centering
\includegraphics[height=3.0in,  width=4in, angle=-90]{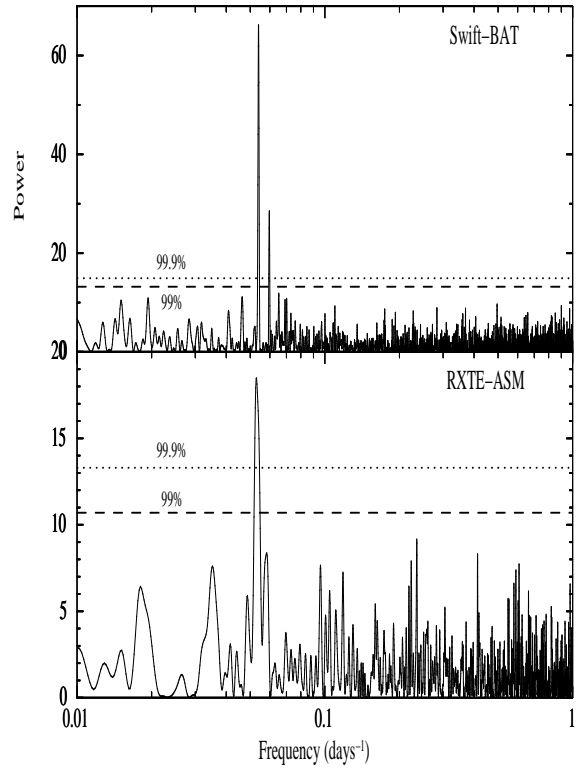}
\caption{The Lomb-Scargle periodogram generated from the $Swift$-BAT and $RXTE$-ASM light curves 
of IGR J18483$-$0311, are respectively shown in the top and the bottom panel. The peak in the top 
panel corresponds to a frequency of 0.0538 d$^{-1}$, .i.e. a period of 18.5873 d. The peak in the 
periodogram generated from the $RXTE$ data corresponds to a frequency of 0.0533 d$^{-1}$, .i.e. a 
period of 18.7617 d. \label{fig:f6}}
\end{figure} 
As shown in Figure~\ref{fig:f6}, the peak in the Lomb-Scargle periodogram of the $Swift$-BAT light 
curve corresponds to 0.0538 d$^{-1}$, .i.e. a period of 18.5873 d. Similarly, from the $RXTE$-ASM 
light curve (Figure~\ref{fig:f6}, bottom panel) a period of 18.7617 d has been found. The dotted and 
the dashed horizontal lines in Figure~\ref{fig:f6} correspond to 99.9$\%$ and 99$\%$ significance 
level as determined from the randomization test explained before. The $Swift$-BAT, $INTEGRAL$-ISGRI 
and $RXTE$-ASM light curves were folded into 32 phasebins with a period of 1602571 s and are shown 
in Figure~\ref{fig:f7}. 
\begin{figure}
\centering
\includegraphics[height=3.0in, width=3.3in, angle=-90]{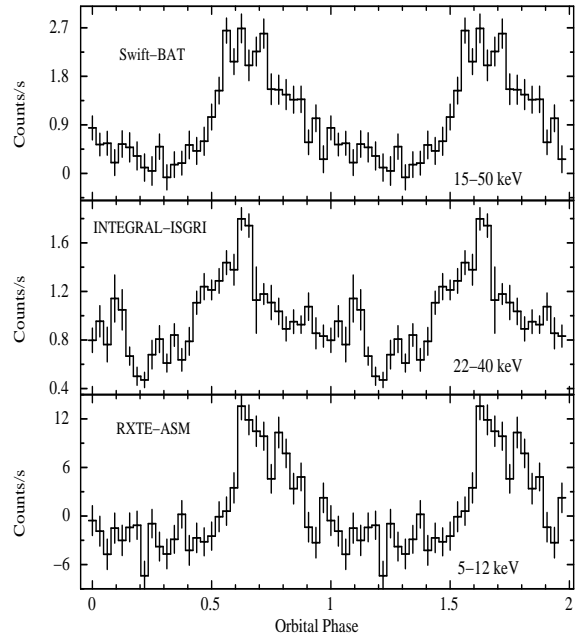}
\caption{The light curve of IGR J18483$-$0311 folded into 32 bins with a period of 1602571 s. The top, 
middle and bottom panels, respectively show the folded light curves of $Swift$-BAT, $INTEGRAL$-ISGRI 
and $RXTE$-ASM observations. \label{fig:f7}}
\end{figure}
Clear peaks are seen in all the three folded light curves.  The folded light curve shows that the source 
is inactive for about half the orbit.

\begin{figure}
\centering
\includegraphics[height=3.0in, width=3.0in, angle=-90]{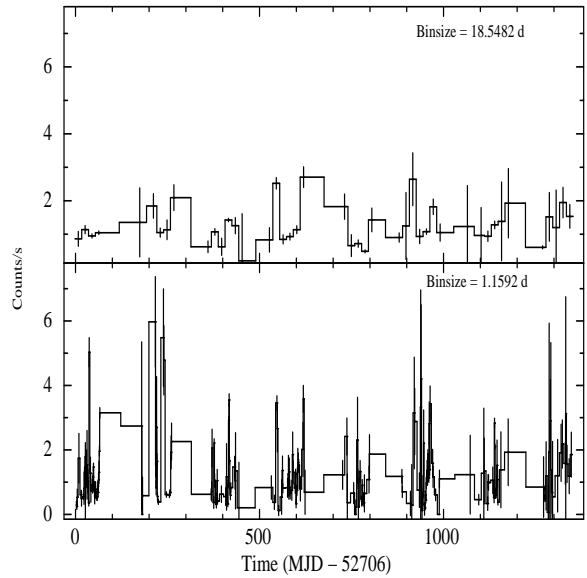}
\caption{The long term light curve of IGR J18483-0311, binned with binsize of 18.5482 d 
and 1.1592 d. \label{fig:f8}}
\end{figure}

\begin{figure}
\centering
\includegraphics[height=3.0in, width=2.5in, angle=-90]{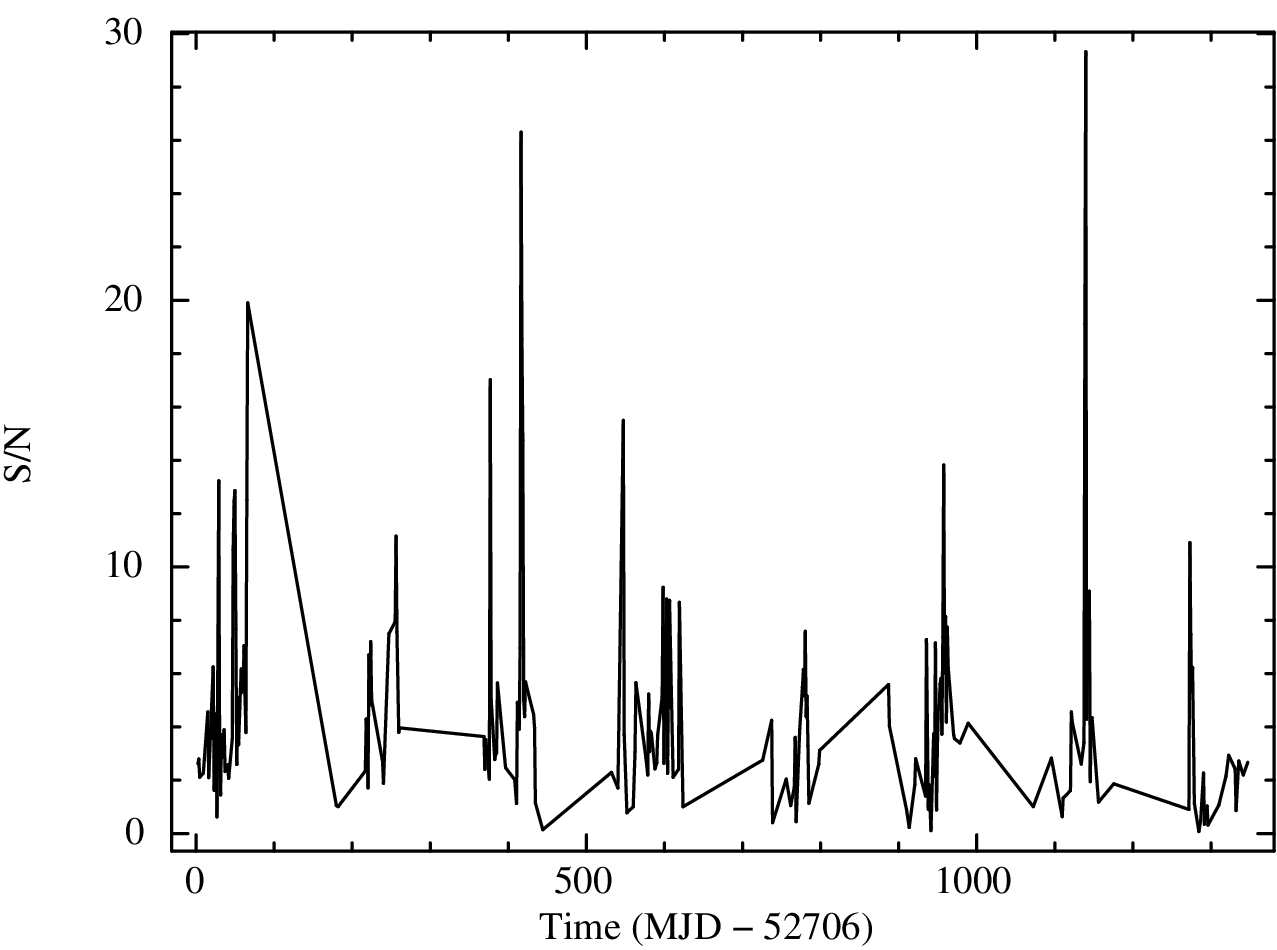}
\includegraphics[height=3.0in, width=2.0in, angle=-90]{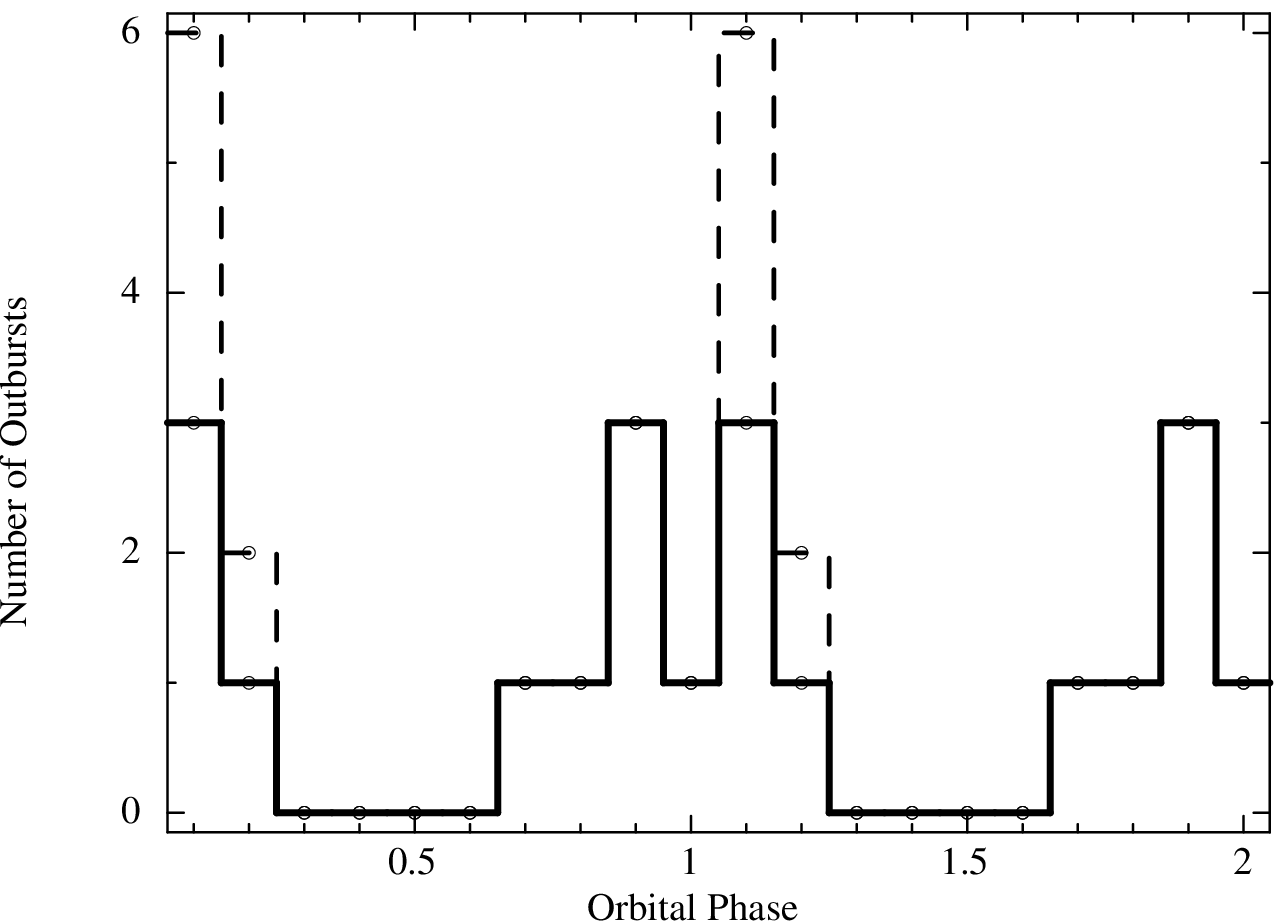}
\caption{The $Swift$-BAT light curve of IGR J18483$-$0311, binned with a binsize of 1.1592 d. The count 
rate was divided by the error. The bottom panel shows the histogram of number of outbursts with respect 
to the orbital phase. The dashed line shows the number of outbursts above the 5 $\sigma$ level and solid 
line shows the number of outbursts above the 10 $\sigma$ level. \label{fig:f9}}
\end{figure}

IGR J18483$-$0311 is a transient source and many outbursts have been recorded by the instruments 
on board $INTEGRAL$. Figure~\ref{fig:f8} shows the long term $INTEGRAL$-ISGRI light curve binned with 
an orbital period of 18.5482 d. The bottom panel of the same figure shows the long term light curve 
binned with a period one-sixteenth of the orbital period. 
Large variations, akin to the SFXT outbursts are seen in the bottom panel compared to the light curve shown 
in the top panel. 
To determine their phase occurence, we took the light curve binned with 1.1592 d and assuming a 
uniform exposure throughout the observation, we divided the signal count rate by the error 
associated with it. The resulting light curve is shown in Figure~\ref{fig:f9} (top panel). We then 
took the outbursts above 5 $\sigma$ and 10 $\sigma$ level and considering an orbital period of 18.5482 d, 
we determined their phase of occurence with respect to the most intense outburst. A histogram of the 
number of outburst in each orbital phase was created. It is shown in the bottom panel 
of Figure~\ref{fig:f9}. The solid curve in the histogram corresponds to outburst above 10 $\sigma$ level 
and the dotted curve is for the 5 $\sigma$ level. As can be seen in the figure, most of the outbursts occur 
at the same phase as the reference $``$most intense" outburst. Although, a few outburst occur at other 
phases also, but this result confirms the periodicity in the occurence of outbursts. 

After the removal of 5$\sigma$ bursts, we searched for an orbital period in the $INTEGRAL$ data and 
the $efsearch$ result is shown in Figure~\ref{fig:f5} (middle panel). The folded $INTEGRAL$ light curve, 
after the removal of 5$\sigma$ outbursts, is shown in Figure~\ref{fig:f7} (middle panel). The profile is 
similar to the folded profile obtained of the $Swift$-BAT and the $RXTE$-ASM light curves.

\begin{figure}
\centering
\includegraphics[height=3.0in, width=4in, angle=-90]{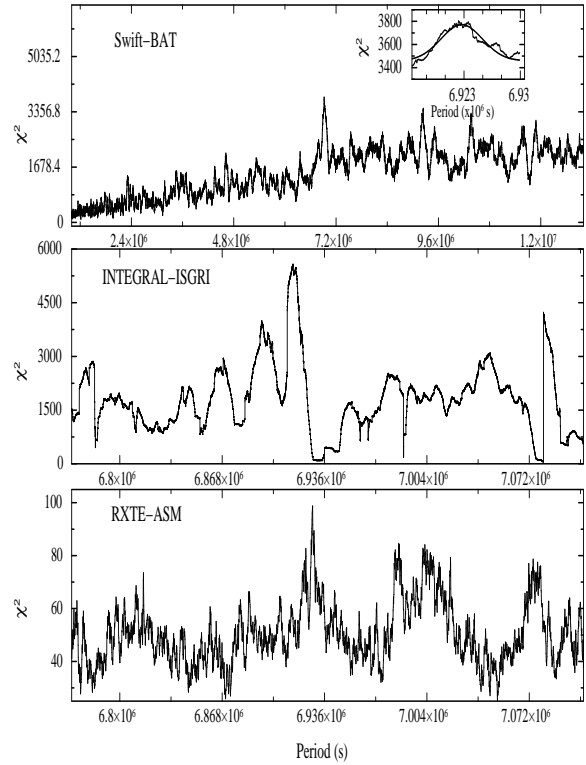}
\caption{Results from $efsearch$ on the light curve of IGR J16318$-$4848. The top panel shows the 
result from the $Swift$-BAT observations, over a wide range of trial periods. The inset figure shows 
the expanded view around the peak determined from the $Swift$-BAT dataset. The solid line represents 
the best fit gaussian curve with the centre at 80.227(14) d. The middle and the bottom panels show 
the $efsearch$ result near the peak from the $INTEGRAL$ and $RXTE$-ASM data. The peaks correspond 
to 80.045(21) d and 80.198(22) d, respectively. \label{fig:f10}}
\end{figure}

\textbf{IGR J16318$-$4848:} As done in the case of IGR J18027$-$2016 and IGR J18483$-$0311, the long 
term $Swift$-BAT, $INTEGRAL$-ISGRI and $RXTE$-ASM light curves were first corrected for the earth 
motion and the periodicity was searched using the ftool $efsearch$. The period search results are 
shown in Figure~\ref{fig:f10}. 
Top panel shows the $efsearch$ result from the $Swift$-BAT light curve and a peak is seen 
near $\sim$ 7 $\times$ 10$^{6}$ s. The period search analysis in IGR J16318$-$4848 is being 
reported for the first time, therefore, we have searched for a period over a wide range of trial 
periods. The inset figure is the expanded view around the peak, about which we fit a gaussian model. 
We obtained the best fit gaussian center as 6931624(1202) s (80.227(14) d). The detection significance 
of periodicity from the $INTEGRAL$ and $RXTE$ observations is very small, but we do detect a peak 
at 6915874(1808) s (80.045(21) d) and 6929084(1912) s (80.198(22) d), respectively. We tried to confirm 
the periodicity using the Lomb-Scargle periodogram technique as done above for the other two sources, 
but we could not detect a significant peak. 

\begin{figure}
\centering
\includegraphics[height=3.in, width=4.0in, angle=-90]{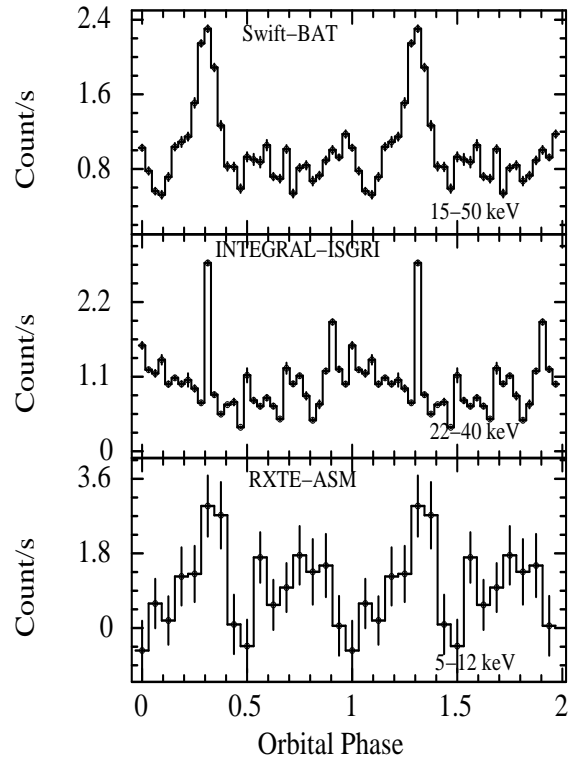}
\caption{The folded $Swift$-BAT, $INTEGRAL$-ISGRI and $RXTE$-ASM light curves of IGR J16318-4848. The 
light curves were folded with the respective best period determined in each case. \label{fig:f11}}
\end{figure}

\begin{figure}
\centering
\includegraphics[height=3.5in, width=3.0in, angle=-90]{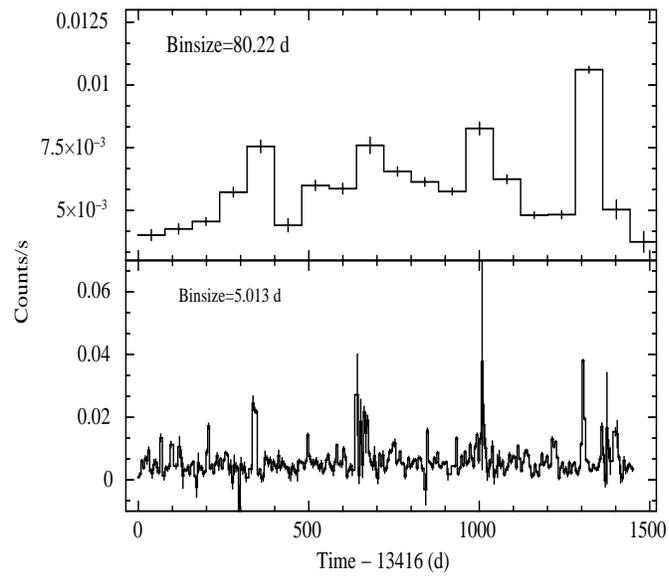}
\caption{The long term light curve of IGR J16318$-$4848, binned with binsize of 80.22 d 
and 5.013 d. \label{fig:f12}}
\end{figure}

Figure~\ref{fig:f11} shows the folded $Swift$-BAT, $INTEGRAL$-ISGRI and $RXTE$-ASM light curves 
of IGR J16318-4848. The light curves were folded with the respective best period determined in each 
case. We have done the rest of the analysis with the $Swift$-BAT light curve, which has the best 
statistics amongst the three observations. 
A clear peak is seen in the folded light curve, along with small secondary peaks. The main peak lasts 
for about 0.2 orbital phase. Figure~\ref{fig:f12} shows the $Swift$-BAT light curve binned with a 
binsize of 80.22 d. The bottom panel of the same figure shows the light curves binned with a binsize 
one-sixteenth of 80.22 d. An intense outburst near the end of the observation clearly stands out and 
the peak observed in the folded orbital light curve could be dominated by this. 

\begin{figure}
\centering
\includegraphics[height=3.0in, width=2.0in, angle=-90]{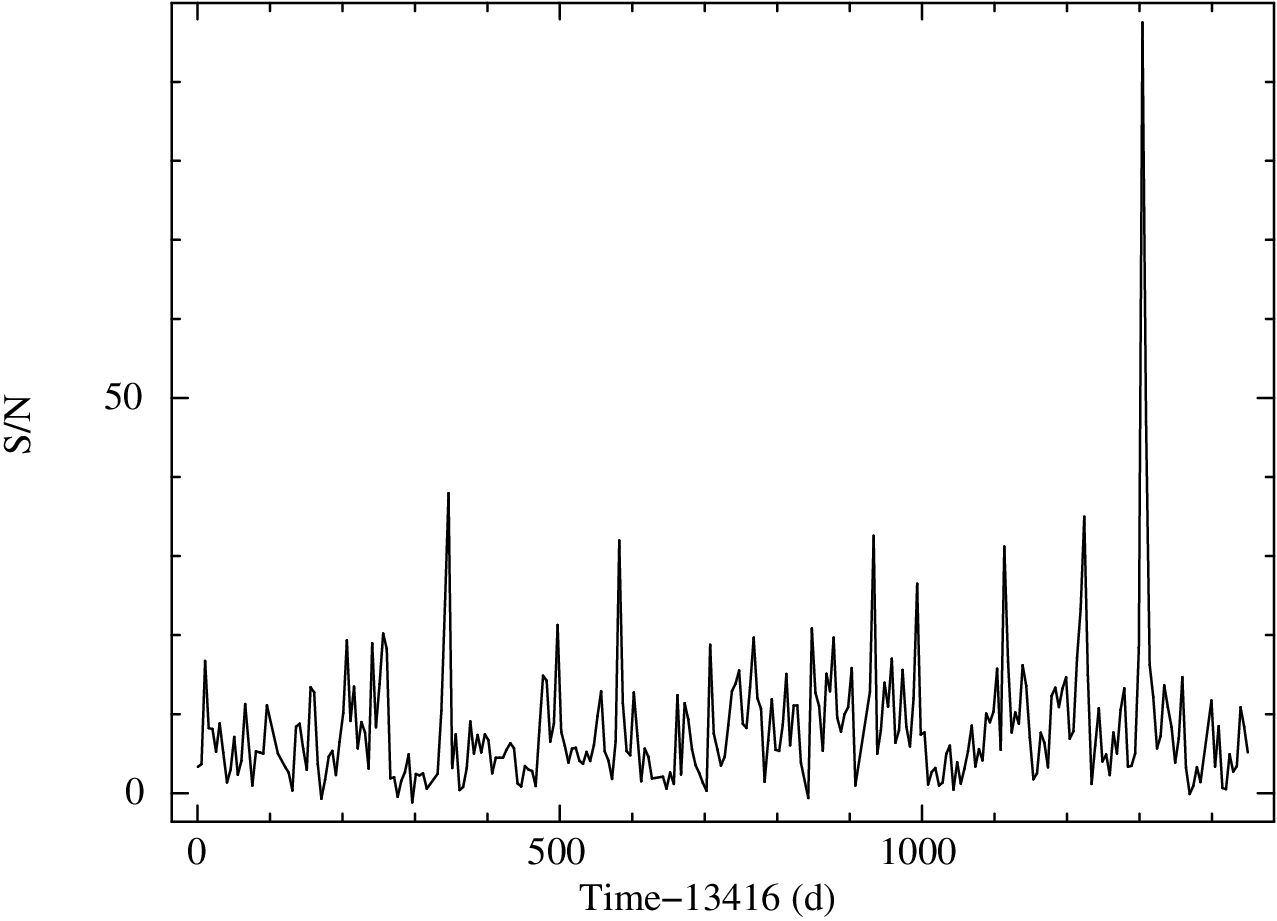}
\includegraphics[height=3.0in, width=2.0in, angle=-90]{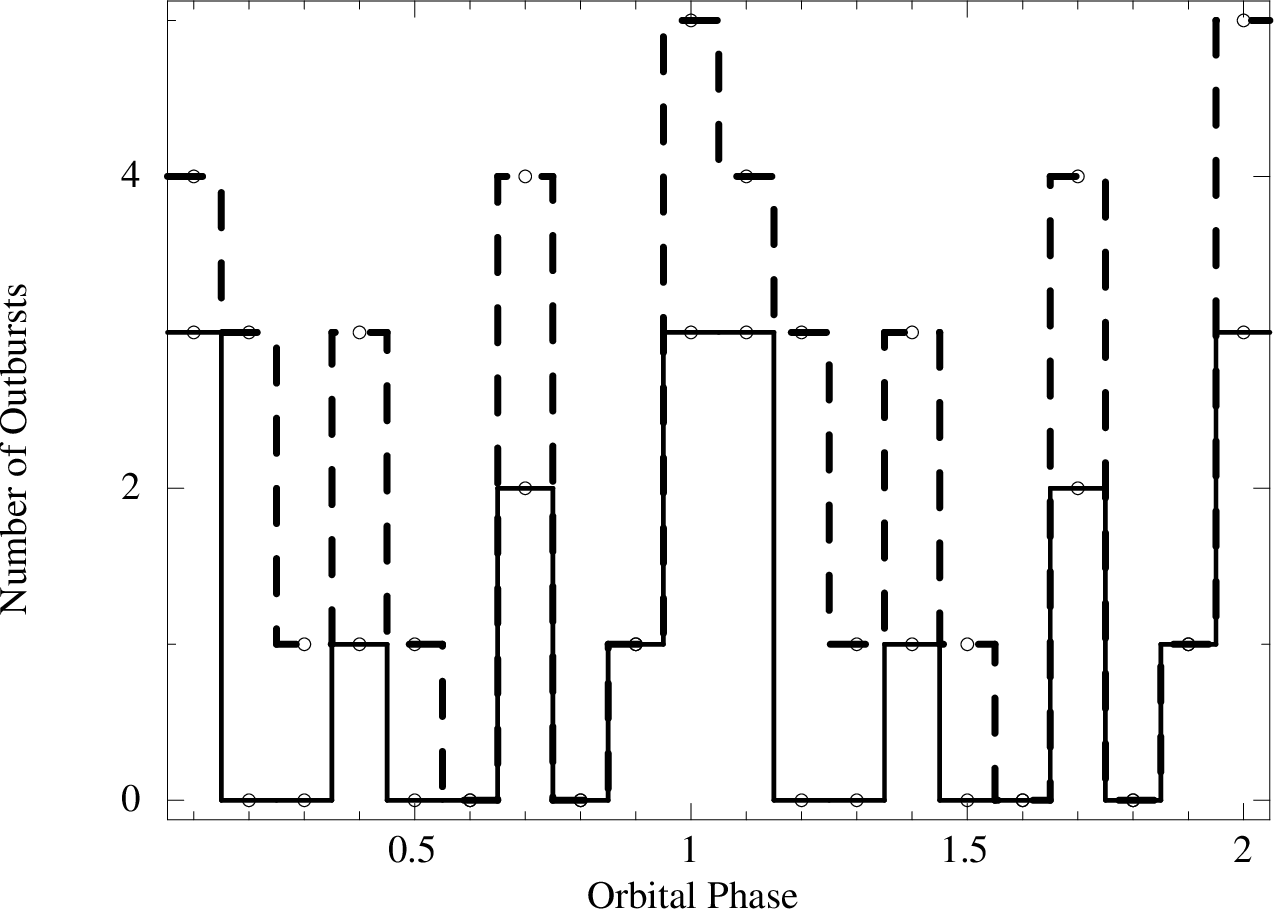}
\caption{The $Swift$-BAT light curve of IGR J16318$-$4848, binned with a binsize of 5.013 d. The count 
rate was divided by the error. The bottom panel shows the histogram of number of outbursts with respect 
to the orbital phase. The dashed line shows the number of outbursts greater than 15 $\sigma$ and solid 
line shows the number of outbursts greater than 20 $\sigma$ level. \label{fig:f13}}
\end{figure}

To check this, we applied a similar analysis as done above for IGR J18483-0311. We divided the signal 
count rate by the error in rate determination for the light curve binned with 5.013 d (shown 
in Figure ~\ref{fig:f13} (top panel)). We then took the outbursts above 15 $\sigma$ and 20 $\sigma$ 
level and considering an orbital period of 80 d, we determined the phase of occurence of outbursts with 
respect to the most intense outburst. The bottom panel of Figure~\ref{fig:f13} shows the histogram of 
number of outbursts in each phase. Most of the outbursts occur around an orbital phase of 0.1, with occasional 
outbursts at phases about 0.4 and 0.7. It implies that though the outbursts occur with a periodicity 
of $\sim$ 80 d, there are three different orbital phases at which they occur. 

\section{Discussion}

Using the long term $Swift$-BAT, $INTEGRAL$-ISGRI and $RXTE$-ASM data of IGR J18027$-$2016, we have 
determined an accurate value of the orbital period of 4.5693(4) d. From the $Swift$-BAT and $RXTE$-ASM 
data, we have accurately determined an orbital period of 18.5482(88) d for IGR J18483$-$0311 and have 
found that the outbursts occur intermittently at intervals of $\sim$18.55 d. We have also found a $\sim$ 80 d 
periodicity in the occurence of outbursts from IGR J16318$-$4848.  

All the three sources, IGR J18027$-$2016, IGR J18483$-$0311 and IGR J16318$-$4848, studied in the present 
work, are bright supergiant High Mass X-ray Binaries which accrete material through the stellar wind of 
a late O/early B-type supergiant companion. The classical supergiant systems have small and circular 
orbits, as compared to relatively larger orbits found in Supergiant Fast X-ray Transients. However, 
there are exceptions to this. Recently, Jain et al. (2009) determined a 3.32 d orbital period for 
the SFXT system IGR J16479$-$4514, which is smaller than that known in 
other SFXTs, IGR J11215$-$5952 (165 d: Romano et al. 2007; Sidoli et al. 2007) 
and SAX J1818.6$-$1703 (30 d: Bird et al. 2009). An orbital period of 4.56 d for IGR J18027$-$2016 is 
well within the expected range for supergiant systems, but an orbital period of 18.5508 d determined 
for IGR J18483$-$0311, is somewhat more than that expected from a supergiant system. IGR J18483$-$0311 
is active for about half the orbital cycle. The quiescent emission level in IGR J18483$-$0311 is also 
higher (Sguera et al. 2007). All this imply that the source is an intermediate system between classical 
supergiants and SFXTs.

IGR J16318$-$4848 is a highly absorbed source with a hydrogen column 
density, N$_{H}$ $\simeq$ 10$^{24}$ cm$^{-2}$. The presence of strong absorption shows that the compact 
object must be embedded in a dense circumstellar envelope, originating from the accretion of stellar 
winds. \textbf{In several HMXBs, the orbital periods have been determined either through the timing 
analysis of the X-ray data (example, IGR J16479$-$4514: Jain et al. (2009); IGR J17544$-$2619: 
Clark et al. (2009); IGR J17252$-$3616: Zurita Heras et al. (2006)) or through the timing of the 
recurrent outbursts (example IGR J11215$-$5952: Sidoli et al. (2007); SAX J1818.6$-$1703: 
Sidoli et al. (2009)).} We have studied the periodicity in the occurence of outbursts in 
this system. The periodicity of $\sim$ 80 d in the outburst behaviour most likely represents the 
orbital period of the binary system and is a key diagnostic for studying the geometry of the system. 

Several models have been proposed to explain the occurrence of periodic outbursts in the supergiant 
systems. In't Zand (2005) suggested the $``$clumpy winds" model according to which the wind from the 
donor star is composed of dense clumps with mass of the order of 10$^{19}$ - 10$^{20}$ g (Howk et al. 2000). 
Neutron star accretes from the wind of the supergiant at different rates depending on the wind density and 
short flares occur due to episodic accretion of clumps from the massive winds. Negueruela et al. (2008) 
suggested that outbursts occur due to accretion of clumps from the spherical wind. They proposed that the 
orbit of these systems are large and the wind clumps density is small. The outbursts in IGR J16318$-$4848 
have been observed to occur at different orbital phases. \textbf{The pattern of the X-ray outbursts depend 
on the size, eccentricity and the orientation of the orbit.} Sidoli et al. (2007) proposed that the supergiant 
wind has an $``$equatorial disk" component, in addition to the spherically symmetric polar component. Outbursts 
occur when the neutron star crosses the equatorial disk component at the periastron, which is denser 
than the polar wind component. The neutron star can cross the disk twice depending on the truncation of 
the disk, its orientation and inclination with respect to the orbital plane. 
 
In view of the results presented above, we point out that more sensitive and frequent monitoring of all 
the three sources is required in order to understand them in detail. Specially in the case 
of IGR J18027$-$2016, which show clear eclipses which can be used to time mark the orbital modulation 
and determine the orbital evolution in the system, if any. Using longer data sets, we have been able 
to determine the orbital period of IGR J18483$-$0311 with greater accuracy. Regular monitoring of the 
absorbed source IGR J16318$-$4848 is important to detect the orbital period with confidence. 
\section*{Acknowledgments}

We thank the $Swift$-BAT and $RXTE$-ASM teams for provision of the data

\end{document}